\documentclass[a4paper, 12pt]{article}
\usepackage{amsmath}
\usepackage{amssymb}
\usepackage{graphicx}
\usepackage{cite}
\def \ni {\noindent}
\def \vs  {\vskip5mm}
\def \bea {\begin{eqnarray}}
\def \cref {c_{\rm{ ref.}}}  
\def \eea {\end{eqnarray}} 
\def \mea {\nonumber\\}
\def \half {{\textstyle \frac{1}{2}}}
\def \xihat {{\hat x}}
\def \mathcalhatC {{\hat c}}

\def \mathcalhatE {{\hat E}}
\def \mathcalhatA {{\hat A}}
\def \hatc {{\hat c}}
\def \Jref {J_{\rm {ref.}}}
\def \Dref {D_{\rm {ref.}}}
\def \Eref {E_{\rm {ref.}}}
\def \hatQ {{\hat Q}}
\def \hatE {{\hat E}}
\def \hatES {{\hatE_{\rm S}}}
\def \hatA {{\hat A}}
\def \hatD {{\hat D}}
\def \hatJ {{\hat J}}
\begin{document}
\begin{titlepage}     
     
\title{Differential equations of electrodiffusion: constant field solutions, uniqueness, and new formulas of Goldman-Hodgkin-Katz type}     
     
\author{
A.J. Bracken\footnote{{\em Email:} a.bracken@uq.edu.au}\, and L. Bass\footnote{{\em Email:} lb@maths.uq.edu.au}\\Department of Mathematics\\    
School of Mathematics and Physics\\University of Queensland\\Brisbane 4072, Australia}   
     
\date{}     
\maketitle     
     
\begin{abstract}
The equations governing one-dimensional, steady-state electrodiffusion
are considered when there are arbitrarily many  mobile ionic species present, in any number of valence classes, possibly also with
a uniform distribution of fixed charges.  Exact constant field solutions and 
new formulas of Goldman-Hodgkin-Katz type are  found. All of these formulas
are exact, unlike the usual approximate ones.
Corresponding boundary conditions on the ionic concentrations are identified.  The
question of uniqueness of constant field solutions with such boundary conditions is considered, and is re-posed 
in terms of an autonomous ordinary differential equation of order $n+1$
for the electric field, where $n$ is the number of valence classes.  
When there are no fixed charges, the equation can be integrated once to give the non-autonomous equation of order $n$ 
considered previously in the literature including, in the case $n=2$, the form of Painlev\'e's second equation considered first
in the context of electrodiffusion 
by one of us. When $n=1$, the new equation is a form of Li\'enard's
 equation.  Uniqueness of the constant field solution is established in this case.  
\end{abstract} 

{\bf Running title:} Differential equations of electrodiffusion


    
\end{titlepage}    
\setcounter{page}{2}
\section{Introduction}
Electrodiffusion --- the transport of charged ions by diffusion and migration --- 
plays a  central role in a variety of important 
physical and biological contexts \cite{teorell1,schwartz,rubinstein}. Ionic species of various valences
may be involved, such as Na$^+$, K$^+$, Cl$^-$, Ca$^{++}$, SO$_4\,^{--}$, $\dots$\,, with various coefficients of diffusion.

The mathematics involved in modeling electrodiffusion  is typically nonlinear, making it a nontrivial task to obtain exact solutions 
or to determine 
the status of
approximations. Of particular interest in the following is the range of validity of so-called `constant field approximations' and 
associated Goldman-Hodgkin-Katz (GHK) formulas \cite{goldman,hodgkin}, notions that have been widely analyzed and extensively employed  
since their introduction,
especially in studies of liquid junctions associated with biological membranes \cite{bass3,moore,zelman1,friedman,agin,zelman2,arndt,jacquez,syganow}.

An extension of the Nernst-Planck steady-state model \cite{nernst,planck}  
incorporates the effect of the
electric field that develops within a liquid junction containing ions in solution,  
with different diffusion coefficients and carrying different charges \cite{grafov,bass1}.  
The extension involves the 
application of Gauss' Law to relate charge density to electric field, together with 
the imposition of
Einstein's relation between diffusion coefficients and mobilities of the ions.  It is the interplay
between diffusion
of the ions at different rates and the action of the ensuing electric 
field on the charges,  that leads to nonlinearity of the model. 

In many cases, the ionic transport can be regarded as one-dimensional, and 
the junction can be modeled as a slab occupying $0\leq x \leq \delta$.  
The ionic species in solution 
can be grouped into $n$ classes with distinct valences
$z_k$, $k=1,\,2,\,\dots\,,n$, with $m_k$ species in the $k$-th valence class.  The $z_k$ and $m_k$ are integers, with $m_k$
positive and $z_k$ positive or negative. 
With  the concentration of the $r$-th species in the 
$k$-th valence class denoted by $c_{kr}(x)$, for $k=1,\,2,\, \dots,\,n$  and  $r=1,\,2,\,\dots\,,m_k$, and with  
the induced electric field within the junction denoted by $E(x)$,
the governing system of coupled first-order ordinary differential equations (ODEs) 
is
\begin{subequations}\label{sys}
\bea
\frac{d c_{kr}(x)}{dx}= \frac{z_k\,e}{k_B \,T}\,E(x)\,c_{kr}(x)+A_{kr}\,,\qquad\qquad
\label{sys_a}
\\ \nonumber
\\ 
\frac{d E(x)}{dx}=\left(\frac{4\pi e}{\epsilon}\right)\,\left[ \sum_{k=1}^n z_k\,\left(\sum_{r=1}^{m_k}c_{kr}(x)
\,\right)+Q\,\right]\,,
\label{sys_b}
\eea
\end{subequations}
for $0<x<\delta$.  Here $A_{kr}=-\Phi_{kr}/D_{kr}$, where $\Phi_{kr}$ denotes the steady 
(constant) flux in the $x$-direction across the junction, of the $r$-th species in the $k$-th valence class, 
and $D_{kr}$ denotes the diffusion coefficient of that species. Furthermore,  $e$ denotes the electronic charge, 
$k_B$ Boltzmann's constant,
$\epsilon$  the electric permittivity, and $T$
the ambient absolute temperature within the solution in the junction.

Equation \eqref{sys_b} allows for the possibility that in
addition to the moving ions there may also be fixed (immobile) ions giving rise to  
a constant and uniform background charge at concentration $Q\,e$.   The necessity to allow for fixed charges in 
naturally occurring 
liquid junctions has long been recognized \cite{meyer,teorell2,goldman}.  They play a key role in the so-called
``glomerular filtration barrier" in the kidney, for example \cite{miner}. 

An important quantity auxiliary to the system of ODEs \eqref{sys} is the 
electric current density
\bea
J=\sum_{k=1}^n z_k\,e\,\left(\sum_{r=1}^{m_k}\Phi_{kr}\,\right)=-\sum_{k=1}^n z_k\,e\,\left(\sum_{r=1}^{m_k}D_{kr}\,A_{kr}
\,\right)
\,.
\label{curr}
\eea 

When supplemented by a prescribed value for $J$ and
suitable prescribed boundary values (BVs) 
at $x=0$ and $x=\delta$ for the concentrations $c_{kr}(x)$,
\bea
c_{kr}(0)=c_{kr0}\,,\quad c_{kr}(\delta)=c_{kr\delta}\,,
\label{concBVs1}
\eea
systems of equations of the form \eqref{sys} specify the 
electrical structure of a wide class of liquid junctions in the steady state, 
including nerve membranes.

These BVs are commonly assumed to be charge-neutral, so that
\bea
\sum_{k=1}^n z_k\,\left(\sum_{r=1}^{m_k}c_{kr0}
\,\right)+Q=0=\sum_{k=1}^n z_k\,\left(\sum_{r=1}^{m_k}c_{kr\delta}
\,\right)+Q\,.
\label{neutrality00}
\eea

An unusual feature of the models described by \eqref{sys} and \eqref{curr} is that the ionic fluxes $\Phi_{kr}$, 
and hence the $A_{kr}$,
although constant, are typically not known {\em a priori} but are among the unknowns to be determined from the 
ODEs, the prescribed BVs \eqref{concBVs1} for the concentrations, and a prescribed value of 
the current density \eqref{curr}.  Thus there are $2(m_1+m_2+\dots +m_n)+1$ unknowns 
(the concentrations, the steady fluxes and the electric field) to be determined from the system of 
$(m_1+m_2+\dots+m_n)+1$ ODEs \eqref{sys}
and the $2(m_1+m_2+\dots +m_n)+1$ boundary conditions (BCs) \eqref{concBVs1}, \eqref{curr}. 
This complicating feature, together with the inherent 
nonlinearity of the ODEs, has impeded mathematical analysis.

Much of that analysis has focused on the case with two species  with valences $+1$ and $-1$, and with $Q=0$, 
which is appropriate for example to junctions in which Na$^+$ (or K$^+$) 
and Cl$^-$ ions move in solution.   
In such a case, with the additional restriction that one of the fluxes vanishes, it was shown 
\cite{grafov} that the concentration corresponding to the vanishing flux satisfies a second order
nonlinear equation 
that can be transformed into the Painlev\'e II ODE (PII) \cite{painleve}.  Independently, it was shown \cite{bass1} 
that in the same  model without this restriction, 
$E(x)$   satisfies another form of PII.  These were perhaps the first appearances of PII in 
physics or biology.  This and other aspects of this simple  model have been 
much studied subsequently (see \cite{rubinstein,ben,zaltzman,rogersA,rogersB,bracken1,bracken2,bass2,macgillivray} and references therein).  
Existence of
solutions for charge-neutral boundary conditions (BCs) 
has been established for this model \cite{thompson,amster}, but it is not known if these solutions are unique.   
Recent analysis \cite{rogersA,rogersB} 
has drawn attention to the relevance to this   model, of 
previously-known exact solutions of PII and of B\"acklund transformations \cite{rogersC} of its 
solutions more generally, leading to the emergent
phenomenon of ``flux quantization" \cite{bracken1,bracken2}.
 
Comparatively little mathematical progress has been made in cases with larger $n$ values, 
in particular on the classical questions of existence and uniqueness of solutions.  The emphasis has been on 
the 
structure of the $n$-th order non-autonomous ODE satisfied by $E(x)$ alone \cite{leuchtag,rogersD}, in 
models with 
$Q=0$ and 
$n\leq 4$, for various values of the $z_k$, but with all $m_k=1$, and with the fluxes regarded as prescribed constants, 
rather than as unknowns to be determined as part of a solution.

In what follows, the general model is considered, 
with any number of valence classes, any number of species in each class,  with fixed charges allowed for, and with 
the constant fluxes regarded
as unknowns to be determined as part of any solution.  

{\em Exact solutions of \eqref{sys} are constructed, with
constant electric field.  The presence of $Q$ in \eqref{sys_b} typically enables a greater richness of such solutions than
would otherwise be the case.  For example, if all $z_k$ have the same sign, it is clear from \eqref{sys_b} that
nontrivial solutions with constant field cannot occur unless $Q$ is nonzero, with the opposite sign. 
In simple cases the exact solutions constructed here
are shown to lead to formulas of GHK type, in a way similar to that by which 
the usual GHK formulas are obtained from {\em approximate} solutions
with constant electric field.  However, there are important differences. 
The formulas obtained here are exact, and most are new, but they  hold under more restrictive BCs 
than  the usual formulas. }

The BCs  appropriate to these exact constant-field solutions and formulas of GHK type 
are identified.  Existence of solutions is not an issue here, as the solutions are exhibited explicitly,
but the question remains as to the uniqueness
of these solutions of the BV problem posed by the system of ODEs with these BCs.

A way to approach this question is suggested, by showing first that   
as a consequence of \eqref{sys}, the field $E(x)$ satisfies an autonomous
ODE of order $(n+1)$, one that has not previously been identified
in the context of electrodiffusion.    When $Q= 0$, this ODE can be integrated once,
leading to the ODE of order $n$ considered previously in the $Q=0$ case \cite{leuchtag,rogersD}.  That known ODE 
depends explicitly on $x$ and on the constant of integration. The simple two-species model described above, 
but now for a general value of  $Q$, 
provides an important example;
when $Q\ne 0$, the ODE satisfied by $E(x)$ is autonomous of order $3$, and when $Q= 0$, this integrates once
to give the familiar $x$-dependent
ODE of order $2$ that is a form of PII.

The BCs to be satisfied by the field and its derivatives up to order $n$ are identified next.  They 
follow from those found
for the system \eqref{sys} as appropriate for constant field solutions.  The uniqueness problem is now reposed 
in terms of the ODE and BCs for the field, and uniqueness is proved in the case that there is just one valence class,
containing any number of species.

\section{Dimensionless formulation}
It is convenient to introduce a suitable reference concentration $\cref$, for example the value of $Q$ or, 
if $Q=0$, 
a typical BV of one of the concentrations.   Then  dimensionless quantities can be defined as
\bea
\xihat=x/\delta\,,\quad \hatc_{kr}(\xihat)=c_k(x)/\cref\,,\quad \hatQ=Q/\cref\,, 
\mea\mea
\hatE(\xihat)=e E(x)\delta/k_B T =E(x)/\Eref\,,\quad \!\!\!{\rm say},
\mea\mea
\hatA_{kr}=A_{kr}\,\delta/\cref\,,\quad \mu=4\pi e^2\delta^2\cref/\epsilon k_B T\,,
\label{dim}
\eea 
enabling  \eqref{sys} to be rewritten in a more transparent, dimensionless form as
\begin{subequations}\label{sys1}
\bea
\hatc_{kr}\,'=z_k\,\hatE\,\hatc_{kr}+\hatA_{kr}\,,\qquad\qquad
\label{sys1_a}
\\ \nonumber
\\
\hatE\,'=\mu\,\left[\sum_{k=1}^n z_k\,\hatc_k\,+\hatQ\right]\,,\qquad\qquad
\label{sys1_b}
\eea
\end{subequations} 
for $k=1,\,2,\,\dots,\,n$ and $r=1,\,2,\,\dots\,m_k$, where
\bea
\hatc_k=\sum_{r=1}^{m_k}\, \hatc_{kr}\,.
\label{ckdef}
\eea
It is also convenient to define
\bea
\hatA_k=\sum_{r=1}^{m_k}\,\hatA_{kr}\,,
\label{Akdef}
\eea
so that 
\bea
\hatc_k\,'=z_k\,\hatE\,\hatc_k +\hatA_k\,.
\label{ckODE}
\eea
In \eqref{sys1} and \eqref{ckODE} the primes denote
differentiation with respect to $\xihat$, and the ODEs are now to hold for $0<\xihat<1$.
The BCs \eqref{concBVs1} become
\begin{subequations}\label{concBVs2}
\bea
\hatc_{kr}(0)=\hatc_{kr0}\,,\quad \hatc_{kr}(1)=\hatc_{kr1}\qquad\qquad\qquad
\label{concBVs2_a}
\\ \nonumber
\\ 
({\rm and \quad \!\!\!\!so}\quad \hatc_k(0)=\sum_{r=1}^{m_k}\,\hatc_{kr0}=\hatc_{k0}\,,\quad \hatc_k(1)=\sum_{r=1}^{m_k}\,\hatc_{kr1}=\hatc_{k1})\,,
\label{concBVs2_b}
\eea
\end{subequations} 
where $\hatc_{kr0}$ and $\hatc_{kr1}$ are positive constants. 
When charge-neutrality is imposed on the boundaries,  \eqref{neutrality00} becomes
\bea
\sum_{k=1}^n\,\left(\sum_{r=1}^{m_k}\, z_k\,\hatc_{kr0}\right)\,+\hatQ \,\,\,(=\sum_{k=1}^n \,z_k\,\hatc_{k0}\,+\hatQ)=0\,,
\mea\mea
\sum_{k=1}^n\,\left(\sum_{r=1}^{m_k}\, z_k\,\hatc_{kr1}\right)\,+\hatQ \,\,\,(=\sum_{k=1}^n \,z_k\,\hatc_{k1}\,+\hatQ)=0\,.
\label{neutrality0}
\eea

The
expression \eqref{curr} for the current density can also be written in dimensionless form
as
\bea 
{\hat J}= J/\Jref=-\sum_{k=1}^n\,z_k \,\left(\sum_{r=1}^{m_k}{\hat D}_{kr}\,
\mathcalhatA_{kr}\right)\,,
\mea\mea
{\rm {where}}\quad \Jref= \left(\frac{e\,\cref\,\Dref}{\delta}\right)\quad {\rm {and}}\quad
{\hat D}_{kr}=D_{kr}/\Dref\,,
\label{curr1}
\eea
with $\Dref$  a suitable reference diffusion coefficient, say the sum of the $D_{kr}$.

\section{Exact constant field solutions and GHK \newline formulas}
The constant $\mu$  defined in \eqref{dim}
is the squared ratio of the junction width $\delta$ to an internal (or bulk) Debye length \cite{agin,macgillivray},
\bea 
\mu=\left(\frac{\delta}{\lambda_{\rm{ Debye}}}\right)^2\,,\quad\lambda_{\rm {Debye}}=\sqrt{\frac{\epsilon k_B T}{4\pi e^2\cref}}\,.
\label{debye}
\eea
The structure of a liquid junction may be such that 
$\mu\ll 1$. Then 
\eqref{sys1_b} reduces to $|\mathcalhatE\,'|\ll 1$, 
implying that
the field is approximately constant throughout the junction.  (Note that there is then no requirement that 
the RHS of \eqref{sys1_b} should vanish, which would require that
charge-neutrality should hold throughout.)
 
With $\hatE$ treated as constant,  \eqref{sys1_a} reduces to a set of uncoupled linear ODEs  for the 
$\mathcalhatC_{kr}(\xihat)$, with constant coefficients.  
These  are  
easily solved using  the BCs \eqref{concBVs2} (compare with \eqref{conc1soln} below),
leading in the case of zero current
density to well-known GHK formulas.        
This is the situation that seems to be  assumed in most 
discussions of those  formulas, 
but the necessity for the condition $\mu\ll 1$ to hold is rarely made explicit \cite{agin,moore}.  Furthermore, uncertain ionic solubilities
in various media lead to uncertainty in the value of $\mu$, and hence in the degree to which the field is approximately constant
\cite{moore}. 

A different approach is taken here. Exact constant-field solutions are identified, and used to derive 
formulas of GHK type.  For such solutions, each side of \eqref{sys1_b} vanishes, 
with no restriction on the size of $\mu$.  
The vanishing of the RHS of \eqref{sys1_b} implies that charge-neutrality holds throughout the junction, including at the boundaries,
that is to say
\bea
\sum_{k=1}^n z_k\,\left(\sum_{r=1}^{m_k}\,\hatc_{kr}(\xihat)\right)+\hatQ
\left(=\sum_{k=1}^n z_k\,\hatc_{k}(\xihat)+\hatQ\right)=0\,,\quad 0< \xihat< 1\,,
\label{neutrality1}
\eea 
in addition to \eqref{neutrality0}.

\subsection{Constant zero field}
Turning then to solutions of \eqref{sys1} with $\hatE(\xihat)=\hatES$, some constant,
consider firstly  $\hatES=0$.   
An associated simple solution of the BV problem posed by \eqref{sys1} and \eqref{concBVs2} can be seen at once to 
be given by
\bea
\hatE(\xihat)=0\,,\quad \hatc_{kr}(\xihat)=\hatc_{kr0}+(\hatc_{kr1}-\hatc_{kr0})\,x\,,\quad 
\hatA_{kr}=\hatc_{kr1}-\hatc_{kr0}\,,
\label{E0soln}
\eea
provided  $\hatc_{kr0}$ and  $\hatc_{kr1}$ satisfy \eqref{neutrality0}.

For this solution,
the current-density is given from \eqref{curr1} and \eqref{E0soln} by
\bea
\hatJ=-\sum_{k=1}^n\,z_k \,\left(\sum_{r=1}^{m_k}{\hat D}_{kr}\,(\hatc_{kr1}-\hatc_{kr0})
\right)\,,
\label{E0curr}
\eea
so \eqref{E0soln} is a solution of the BV problem posed by \eqref{sys1}, \eqref{concBVs2}, \eqref{neutrality0} and \eqref{E0curr}.
Even in this simple case, it is not clear for general $n$ and $m_k$ if this is the unique solution of that BV problem.
Note that when $n=1$ and $m_1=1$, \eqref{E0soln} collapses to a solution with $\hatE(\xihat)=0$, 
$\hatc_{11}(\xihat)=-\hatQ/z_1$, $\hatA_{11}=0$ and $\hatJ=0$.

\subsection{Constant nonzero field}
Consider next constant-field solutions of \eqref{sys1} with $\hatES\ne 0$. 
Now \eqref{ckODE} becomes
\bea
\hatc_k\,'=z_k\,\hatES\,\hatc_k+\hatA_k\,,
\label{ckODE2}
\eea
with general solution
\bea
\hatc_k(\xihat)=-\frac{\hatA_k}{z_k\,\hatES}+\gamma_k\,e^{z_k\,\hatES\,\xihat}\,,
\label{cksoln}
\eea
where the $\gamma_k$ are arbitrary constants.   Because the exponentials here with different $z_k$ are linearly
independent, \eqref{neutrality1} requires that all the $\gamma_k$ vanish.  Thus  $\hatc_k(\xihat)$ is constant
for each value of $k$,
\bea
\hatc_k(\xihat)= -\frac{\hatA_k}{z_k\,\hatES}=\hatc_k^*\!\!\quad{\rm say},\!\!\quad 0\le \xihat\le 1\,,
\label{cksoln1}
\eea
and \eqref{neutrality1} then requires that
\bea
\sum_{k=1}^n z_k\,\hatc_k^*=-\hatQ\,.
\label{cksoln2}
\eea
From \eqref{cksoln1} it now follows that
\bea
\sum_{k=1}^n \hatA_k=\hatES\,\hatQ\,,
\label{Ahatcondn}
\eea
and that
\bea
\sum_{r=1}^{m_k}\,\hatc_{kr}(\xihat)=\sum_{r=1}^{m_k}\,\hatc_{kr0}=\sum_{r=1}^{m_k}\,\hatc_{kr1}=
\hatc_k^*\,.
\label{ckrsoln}
\eea

For
any $k$ value such that $m_k=1$, it follows from \eqref{cksoln1} and \eqref{ckrsoln} that
\bea
\hatc_{k1}(\xihat)=\hatc_k^*\,,\quad \hatA_{k1}=-z_k\,\hatc_k^*\,\hatES\,,
\label{ck1soln}
\eea
and for 
any $k$ value such that $m_k>1$, it follows 
from \eqref{sys1_a} and the first two of \eqref{concBVs2}    that
\bea
\hatc_{kr}(\xihat)=-\frac{\hatA_{kr}}{z_k\,\hatES}+\left(\hatc_{kr0}+
\frac{\hatA_{kr}}{z_k\,\hatES}\right)\,e^{z_k\,\hatES\,\xihat}\,,
\label{conc1soln}
\eea
together with the condition
\bea
\hatc_{kr1}=-\frac{\hatA_{kr}}{z_k\,\hatES}+\left(\hatc_{kr0}+
\frac{\hatA_{kr}}{z_k\,\hatES}\right)\,e^{z_k\,\hatES}\,,
\label{conc1constraint}
\eea
or equivalently
\bea
\hatA_{kr}
=(\hatc_{kr0}\,e^{z_k\,\hatES}-\hatc_{kr1})\,\left(\frac{z_k\,\hatES}{1-e^{z_k\,\hatES}}\right)\,.
\label{conc1constraint2}
\eea
Note with the help of \eqref{Ahatcondn} that \eqref{conc1soln} is consistent with the global charge-neutrality condition
\eqref{neutrality1} and the condition \eqref{ckrsoln}.

Multiplying $\hatA_{kr}$ by $z_k\,\hatD_{kr}$ using \eqref{ck1soln} or \eqref{conc1constraint2}
as appropriate,  summing over $k$ and $r$ values, then using 
\eqref{curr1}, leads to a formula expressing $\hatJ$ in terms of $\hatES$ and the boundary values \eqref{concBVs2},
which in principle determines $\hatES$ in terms of the BCs.

\subsubsection{Special cases}

\ni
$\bullet \quad{\bf n}$ {\bf arbitrary,} ${\bf m_1=m_2=\dots =m_n=1}$

Here each $\hatA_{k1}$ is as in  \eqref{ck1soln}, leading from \eqref{curr1} to
\bea
\hatJ=\sum_{k=1}^n\,z_k^2\,\hatc_k^*\,\hatES\,,\quad
{\rm so\quad that}\quad
\hatES=\frac{\hatJ}{\sum_{k=1}^n\,z_k^2\,\hatc_k^*}\,.
\label{JEeqn1}
\eea
Then \eqref{ck1soln} gives
\bea
\hatc_{k1}=\hatc_k^*\,,\quad \hatA_{k1}=- \frac{z_k\,\hatc_k^*\,\hatJ}{\sum_{k=1}^n\,z_k^2\,\hatc_k^*}\,,
\label{fullsoln1}
\eea
completing a solution of the BV problem posed by \eqref{sys1} with the BCs \eqref{curr1} and
\bea
\hatc_{k10}=\hatc_{k11}=\hatc_k^*\,.
\label{fullsoln2}
\eea

\ni
$\bullet\quad{\bf n}$ {\bf arbitrary, all} ${\bf m_k>1}$

Each $\hatA_{kr}$ is now given by \eqref{conc1constraint2}, leading to  
\bea
\hatJ=-\sum_{k=1}^n\,\frac{z_k^2\,\hatES}{e^{z_k\,\hatES}-1}\left(\,\sum_{r=1}^{m_k}\,\hatD_{kr}\,\left(\hatc_{kr1}
-\hatc_{kr0}\,e^{z_k\,\hatES}\right)\right)\,.
\label{JEeqn2}
\eea
This transcendental equation cannot be solved explicitly, but  determines $\hatES$ implicitly
after  $\hatJ$ is prescribed, together with prescribed boundary values \eqref{concBVs2} 
for the concentrations, satisfying
the constraints \eqref{cksoln2},  \eqref{ckrsoln}. 

Once $\hatES$ is determined, if only implicitly,  
then $\hatA_{kr}$ is given by \eqref{conc1constraint2}, and $\hatc_{kr}(\xihat)$ is given by
\eqref{conc1soln}, and a solution of \eqref{sys1} with that value for $\hatJ$ and those boundary values is complete.

When $\hatJ=0$, \eqref{JEeqn2} simplifies 
to
\bea
\sum_{k=1}^n\,\frac{z_k^2}{e^{z_k\,\hatES}-1}\,\sum_{r=1}^{m_k}\,\hatD_{kr}\,\hatc_{kr0}\,e^{z_k\,\hatES}
= 
\sum_{k=1}^n\,\frac{z_k^2}{e^{z_k\,\hatES}-1}\,\sum_{r=1}^{m_k}\,\hatD_{kr}\,\hatc_{kr1}\,.
\label{JEeqn3}
\eea
This is still intractable in general, but can be solved for $\hatES$ in the following  special cases
which, though simple mathematically, are nevertheless important for applications.

\vs\ni
$\bullet\quad{\bf n=1,\,m_1>1,\,\hatJ=0}$

Here \eqref{JEeqn3} simplifies further to 
\bea
e^{z_1\,\hatES}\,\sum_{r=1}^{m_1}\,\hatD_{1r}\,\hatc_{1r0}=\sum_{r=1}^{m_1}\,\hatD_{1r}\,\hatc_{1r1}
\label{JEequn4}
\eea
leading to the formula
\bea
\hatES=\frac{1}{z_1}\,\log\left(\frac{\sum_{r=1}^{m_1}\,\hatD_{1r}\,\hatc_{1r1}}
{\sum_{r=1}^{m_1}\,\hatD_{1r}\, \hatc_{1r0}}\right)\,.
\label{GHK1}
\eea
In terms of the original, dimensional variables, this reads
\bea
E_{\rm{S}}=\frac{k_B T}{z_1\,e\delta}\,\log\left(\frac{\sum_{r=1}^{m_1}\,D_{1r}\,c_{1r}(\delta)}{\sum_{r=1}^{m_1}\,D_{1r}\, c_{1r}(0)}\right)\,,
\label{GHK2}
\eea
while the charge-neutrality condition \eqref{neutrality1} becomes
\bea
Q+z_1\,\sum_{r=1}^{m_1}\,c_{1r}(x)=0\,,\quad 0\leq x\leq \delta\,.
\label{neutrality4}
\eea

With $\hatES$ determined, explicit formulas for $\hatA_{1r}$ and
$\hatc_{1r}(\xihat)$, and hence for $A_{1r}$ and $c_{1r}(x)$, can now be constructed if desired from
\eqref{conc1constraint2} and \eqref{conc1soln} to complete
the solution of \eqref{sys1} and hence of \eqref{sys}. 

\vs\ni
$\bullet\quad{\bf n=2,\,z_2=-z_1,\,m_1>1}$ {\bf and} ${\bf m_2>1,\,\hatJ=0}$ 

Here \eqref{JEeqn3} simplifies to
\bea
\frac{1}{e^{z_1\,\hatES}-1}\,\sum_{r=1}^{m_1}\,\hatD_{1r}\,\left(\hatc_{1r1}-\hatc_{1r0}\,e^{z_1\,\hatES}\right)\qquad\qquad\qquad
\mea\mea
=
\frac{-1}{e^{-z_1\,\hatES}-1}\,\sum_{r=1}^{m_2}\,\hatD_{2r}\,\left(\hatc_{2r1}-\hatc_{2r0}\,e^{-z_1\,\hatES}\right)
\label{GHK3}
\eea
and hence to
\bea
\sum_{r=1}^{m_1}\,\hatD_{1r}\,\left(\hatc_{1r1}-\hatc_{1r0}\,e^{z_1\,\hatES}\right)
=
\sum_{r=1}^{m_2}\,\hatD_{2r}\,\left(\hatc_{2r1}\,e^{z_1\,\hatES}-\hatc_{2r0}\right)
\label{GHK4}
\eea
and then
\bea
e^{z_1\,\hatES}\, \left(\sum_{r=1}^{m_1}\,\hatD_{1r}\,\hatc_{1r0}+\sum_{r=1}^{m_2}\,\hatD_{2r}\,\hatc_{2r1}\right) 
=
\left(\sum_{r=1}^{m_1}\,\hatD_{1r}\,\hatc_{1r1}+\sum_{r=1}^{m_2}\,\hatD_{2r}\,\hatc_{2r0}\right)\,,
\label{GHK5}
\eea
so that
\bea
\hatES=\frac{1}{z_1}\,\log\left(\frac{\sum_{r=1}^{m_1}\,\hatD_{1r}\,\hatc_{1r1}+\sum_{r=1}^{m_2}\,\hatD_{2r}\,\hatc_{2r0}}
{\sum_{r=1}^{m_1}\,\hatD_{1r}\,\hatc_{1r0}+\sum_{r=1}^{m_2}\,\hatD_{2r}\,\hatc_{2r1}}\right)
\label{GHK6}
\eea
and, in dimensional form,
\bea
E_{\rm{S}}=\frac{k_B T}{z_1\,e\delta}\,\log\left(\frac{\sum_{r=1}^{m_1}\,D_{1r}\,c_{1r}(\delta)+\sum_{r=1}^{m_2}\,D_{2r}\,c_{2r}(0)}
{\sum_{r=1}^{m_1}\,D_{1r}\,c_{1r}(0)+\sum_{r=1}^{m_2}\,D_{2r}\,c_{2r}(\delta)}\right)\,.
\label{GHK7}
\eea

Two further cases with $\hatJ=0$ 
that involve $\hatA_{kr}$ of the forms in both \eqref{ck1soln} and \eqref{conc1constraint2} are  of interest.

\vs\ni
$\bullet\quad{\bf n>1}$ {\bf and} ${\bf m_1>1,\,}$ ${\bf m_2=m_3=\dots =m_n=1,\,\hatJ=0}$

Here $\hatA_{1r}$ is as in \eqref{conc1constraint2} while $\hatA_{kr}$ is as in \eqref{ck1soln} for $k>1$, leading 
with $\hatJ=0$ to
\bea
\frac{z_1^2}{1-e^{z_1\,\hatES}}\,\sum_{r=1}^{m_1}\,\hatD_{1r}\,\left(\hatc_{1r1}-\hatc_{1r0}\,e^{z_1\,\hatES}\right)
=\sum_{k=2}^n\,z_k^2\,D_{k1}\,\hatc_k^*\,,
\label{GHK8}
\eea
and hence, in dimensional form,  to
\bea
E_{\rm{S}}=\frac{k_B T}{z_1\,e\delta}\,\log\left(\frac{\sum_{r=1}^{m_1}\,z_1^2\,D_{1r}\,c_{1r}(\delta)+\sum_{k=2}^n\,z_k^2\,D_{k1}\,c_k^*}
{\sum_{r=1}^{m_1}\,z_1^2\,D_{1r}\,c_{1r}(0)+\sum_{k=2}^n\,z_k^2\,D_{k1}\,c_k^*}\right)\,,
\label{GHK9}
\eea
where $c_k(0)=c_k(\delta)=c_k^*$.

\vs\ni
$\bullet\quad{\bf n>2,\,}$  ${\bf z_2=-z_1,\,m_1>1,\,m_2>1,\, m_3=m_4=\dots =m_n=1,\,\hatJ=0}$

Now $\hatA_{1r}$ and $\hatA_{2r}$ are as in \eqref{conc1constraint2}, with opposite values of $z_k$, 
while $\hatA_{kr}$ is as in \eqref{ck1soln} for $k>2$.  With 
$\hatJ=0$, this leads to 
\bea
\frac{z_1^2}{1-e^{z_1\,\hatES}}\,\sum_{r=1}^{m_1}\,\hatD_{1r}\,\left(\hatc_{1r1}-\hatc_{1r0}\,e^{z_1\,\hatE_S}\right)
\qquad\qquad\qquad\qquad\qquad\qquad
\mea\mea
+
\frac{z_1^2}{1-e^{-z_1\,\hatE_S}}\,\sum_{r=1}^{m_2}\,\hatD_{2r}\,\left(\hatc_{2r1}-\hatc_{2r0}\,e^{-z_1\,\hatES}\right)
+\sum_{k=3}^n\,z_k^2\,\hatD_{k1}\hatc_k^*=0\,,
\label{GHK10}
\eea
and hence to 
\bea
E_{\rm{S}}=\frac{k_B T}{z_1\,e\delta}\,\times
\qquad\qquad\qquad\qquad\qquad\qquad\qquad\qquad\qquad
\mea\mea
\log \left(\frac{\sum_{r=1}^{m_1}\,z_1^2\,D_{1r}\,c_{1r}(\delta)
+\sum_{r=1}^{m_2}\,z_1^2\,D_{2r}\,c_{2r}(0)
+\sum_{k=3}^n\,z_k^2\,D_{k1}\,c_k^*}
{\sum_{r=1}^{m_1}\,z_1^2\,D_{1r}\,c_{1r}(0)+\sum_{r=1}^{m_2}\,z_1^2\,D_{2r}\,c_{2r}(\delta)+\sum_{k=2}^n\,z_k^2\,D_{k1}\,c_k^*}\right)\,.
\label{GHK11}
\eea

While the formulas \eqref{JEeqn1}, \eqref{GHK2}, \eqref{GHK9} and \eqref{GHK11} appear to be new, 
\eqref{GHK7} has the same form as the original GHK formula \cite{goldman,hodgkin} for the 
corresponding set of ionic species.  But it must 
be emphasized that the basis for the derivation given above is  different in {\em all} these cases, arising not from  an approximation
dependent upon a very small ratio of the membrane width $\delta$ to the internal 
Debye length $\lambda_{\rm Debye}$ of \eqref{debye}, 
but rather from an exact result 
that is a  consequence of more restrictive BCs as in \eqref{neutrality0} and the possible presence of fixed charges
leading to charge-neutrality throughout.  This is an important distinction to be borne in mind, even when
comparing \eqref{GHK7} and the original GHK formula with the same form. 

The distinction is illustrated by the following  two numerical studies relating to formula \eqref{GHK7} for 
a hypothetical junction containing four ionic species, two with $z=+1$ and two with $z=-1$, 
taken here to be Chol$^+$ (choline), K$^+$ (potassium), Cl$^-$ (chlorine) and 
Prop$^-$ (propionate). 
The corresponding diffusion coefficients in dimensionless form  were calculated as in \eqref{curr1} from experimental
values \cite{unsw}, as ${\hat D}_{\rm Chol}=0.17$, ${\hat D}_{\rm K}=0.33$, ${\hat D}_{\rm Cl}=0.34$ and ${\hat D}_{\rm Prop}=0.16$. 
Boundary values of the concentrations, relative to an arbitrary reference concentration $c_{\rm ref.}$,
were chosen to be $0.2$ and $2.5$ for choline (at ${\hat x}=0$ and ${\hat x}=1$, respectively), $2.4$ and $0.1$ for potassium, $0.5$ 
and $3.4$ for chlorine, and $3.1$ and $0.2$ for propionate.  Then the sum of the choline and potassium concentrations is the same at the two faces, 
as is 
the sum of the chlorine and propionate concentrations, consistent with the conditions on BVs in \eqref{ckrsoln}.  The current-density was set to ${\hat J}=0$.

Fig. 1 relates to the situation where the junction also contains 
fixed charges at a positive concentration $Q/\cref=1.0$, so that the
charge-neutrality conditions \eqref{neutrality0} also hold at each boundary face.   In this situation all the conditions necessary for the existence of
an exact constant-field solution are satisfied, in which case formula \eqref{GHK6} and hence \eqref{GHK7} holds exactly, as derived above.   
The BV problem for the system of 
ODEs \eqref{sys1} in this case was solved using MATLAB \cite{matlab}, 
leading to the plots in Fig. 1. The electric field was indeed
found to be constant, with non-dimensionalised value 
$\hatES=-0.58$, which checks with the value given by the formula \eqref{GHK6}.  
Corresponding values of the non-dimensionalised ionic fluxes  were also obtained in the process,  as
${\hat A}_{\rm Chol}=3.15$, ${\hat A}_{\rm K}=-1.64$, ${\hat A}_{\rm Cl}=1.84$, ${\hat A}_{\rm Prop}=-3.94$.

\begin{figure}[ht]
\centering
\includegraphics[width=5in]
{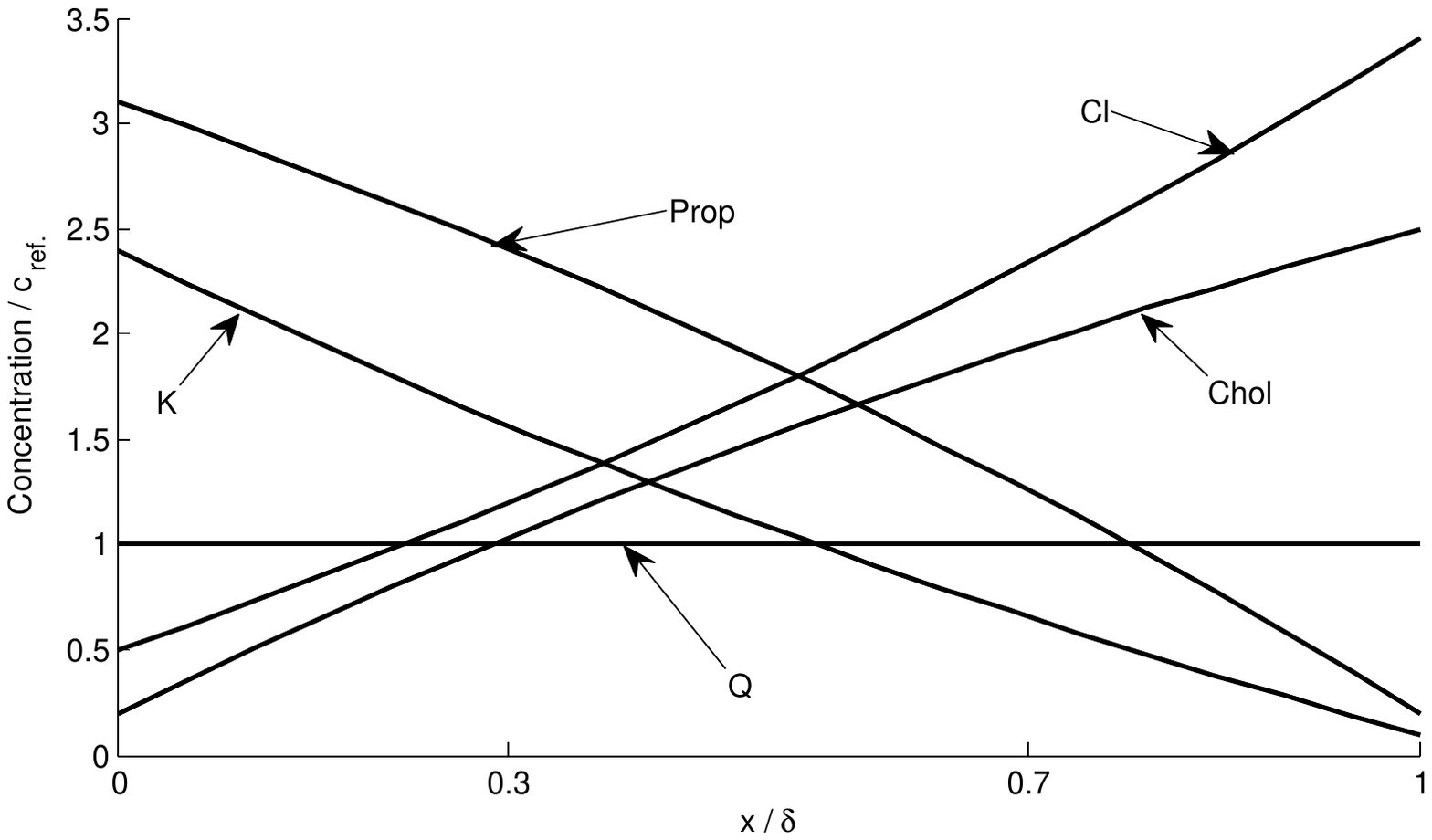}
\caption{Concentration plots.
Fixed charges ensure global charge-neutrality and exact constant field. }
\end{figure}

Fig. 2 relates to the constant-field approximation for the same system, with the same BCs, and again with ${\hat J}=0$,
but now with no fixed charges, so that $Q=0$. 
The plots on the left show the concentration curves for three values of $\mu$, while those on the right show the corresponding 
graphs of the electric field.  They show clearly the nature of the usual approximate approach,
with the concentration curves approaching those of Fig.1 as $\mu$ decreases in size, and the electric field approaching, 
but never equal to  the constant value $-0.58$ given by the GHK formula \eqref{GHK6}, which is not exact in this approximate treatment.   

\begin{figure}[ht]
\centering
\includegraphics[width=5in]
{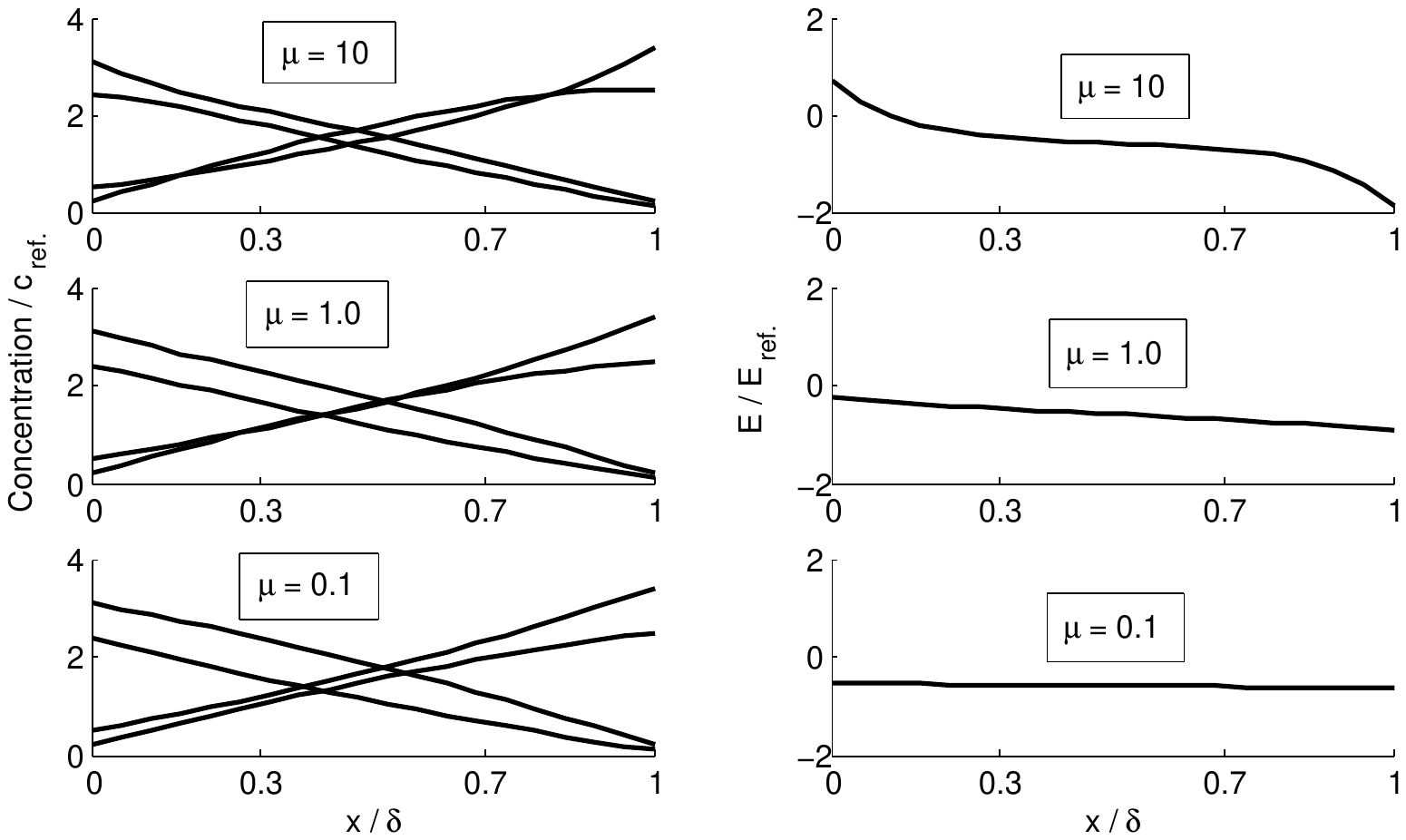}
\caption{Concentration and electric field plots. No fixed charges, no charge-neutrality, constant-field approximation improves as $\mu$ decreases.}
\end{figure}

It is remarkable that none of the exact formulas \eqref{GHK2}, \eqref{GHK7}, \eqref{GHK9} and \eqref{GHK11} 
for $\hatES$ involves the values of $\mu$ or $\hatQ$ explicitly.  
Indeed, the value of $\mu$ has no influence on the value of $\hatES$ in these formulas 
(nor on the plots in Fig. 1, for example),
in sharp contrast to the usual constant-field approximation, which improves as the value of 
$\mu$ decreases towards zero. On the other hand,  
the value of $\hatQ$ is involved   implicitly in the exact formulas through the charge-neutrality
conditions \eqref{neutrality1}
and \eqref{neutrality0}.  For example, in 
the cases associated with \eqref{GHK2} the condition \eqref{neutrality4}, which is implied
in the derivation of \eqref{GHK2},
can only be satisfied if  $\hatQ$ is nonzero, with opposite sign to $z_1$. 

The exact formulas  \eqref{GHK2}, \eqref{GHK7}, \eqref{GHK9} and \eqref{GHK11} were obtained above 
from \eqref{JEeqn2}, for particular ionic configurations corresponding to special values of $n$ and the $m_r$, 
and with in addition the assumption that 
$\hatJ=0$.  In other cases, \eqref{JEeqn2} and \eqref{JEeqn3} 
must be solved numerically to obtain $\hatES$ for 
given $\hatJ$.  This is done most easily by determining $\hatJ$ values directly from \eqref{JEeqn2}
for a sufficient density of $\hatES$ values,
and then plotting those $\hatES$ values against the corresponding  
$\hatJ$ values.  Values of the $\hatA_{kr}$ at corresponding points can be 
obtained using \eqref{conc1constraint2} and whatever concentration boundary values are prescribed, and then also plotted
against $\hatJ$ values.  Once $\hatES$ and the $\hatA_{kr}$ are effectively determined as functions of $\hatJ$
in this way, ionic concentrations can be determined from \eqref{conc1soln} and plotted against $\xihat$.

Fig. 3 shows plots of $\hatES$, $\hatA_{11}$ and $\hatA_{12}$ {\em v.} $\hatJ$ obtained in this way from \eqref{JEeqn2}
in the case $n=1$, $z_1=1$, $m_1=2$, with diffusion coefficients taken to be those for K$^{+}$ 
and Chol$^{+}$ at non-dimensionalised concentrations $\hatc_{11}(\xihat)$ and $\hatc_{12}(\xihat)$ respectively.
The BCs were taken to be $\hatc_{110}=0.25$, $\hatc_{111}=0.55$, $\hatc_{120}=0.75$, $\hatc_{121}=0.45$, with 
$\hatQ=-1$ to ensure charge-neutrality.  
As checks on the numerical calculations, 
the circled points on the plots in Fig. 3 are at the `exact' values obtained  from \eqref{GHK1} when $\hatJ=0$.

\begin{figure}[ht]
\centering
\includegraphics[width=5in,angle=90]
{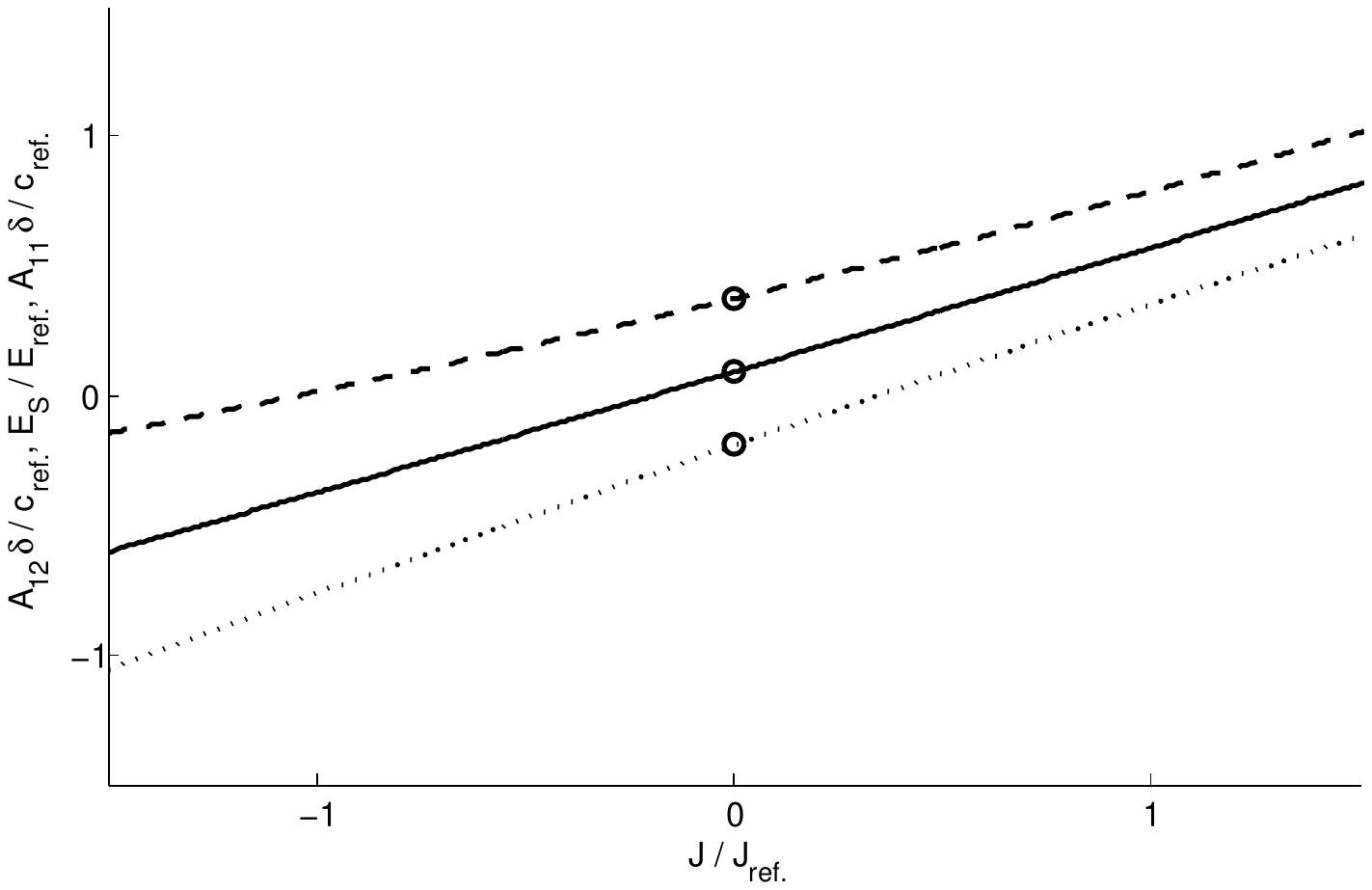}
\caption{Plots of $\hatES$ (solid), $\hatA_{11}$ (dashed) and $\hatA_{12}$ (dotted) {\em v.} $\hatJ$.  Circled points
 at $\hatJ=0$ obtained independently. }
\end{figure}

Fig. 4 shows plots of the corresponding 
concentrations $\hatc_{11}$ (or K$^+$) and $\hatc_{12}$ (or Chol$^+$) {\em v.} $\xihat$ when
$\hatJ=1$ (dashed), $\hatJ=0$ (solid) and $\hatJ=-1$ (dotted).  Upper curves are for $\hatc_{11}$, lower curves for
$\hatc_{12}$.  For each value of $\hatJ$,  charge-neutrality can be seen to hold throughout the junction, with
$\hatc_{11}+\hatc_{12}+\hatQ=0$ for all $\xihat$.  

\begin{figure}[ht]
\centering
\includegraphics[width=5in]
{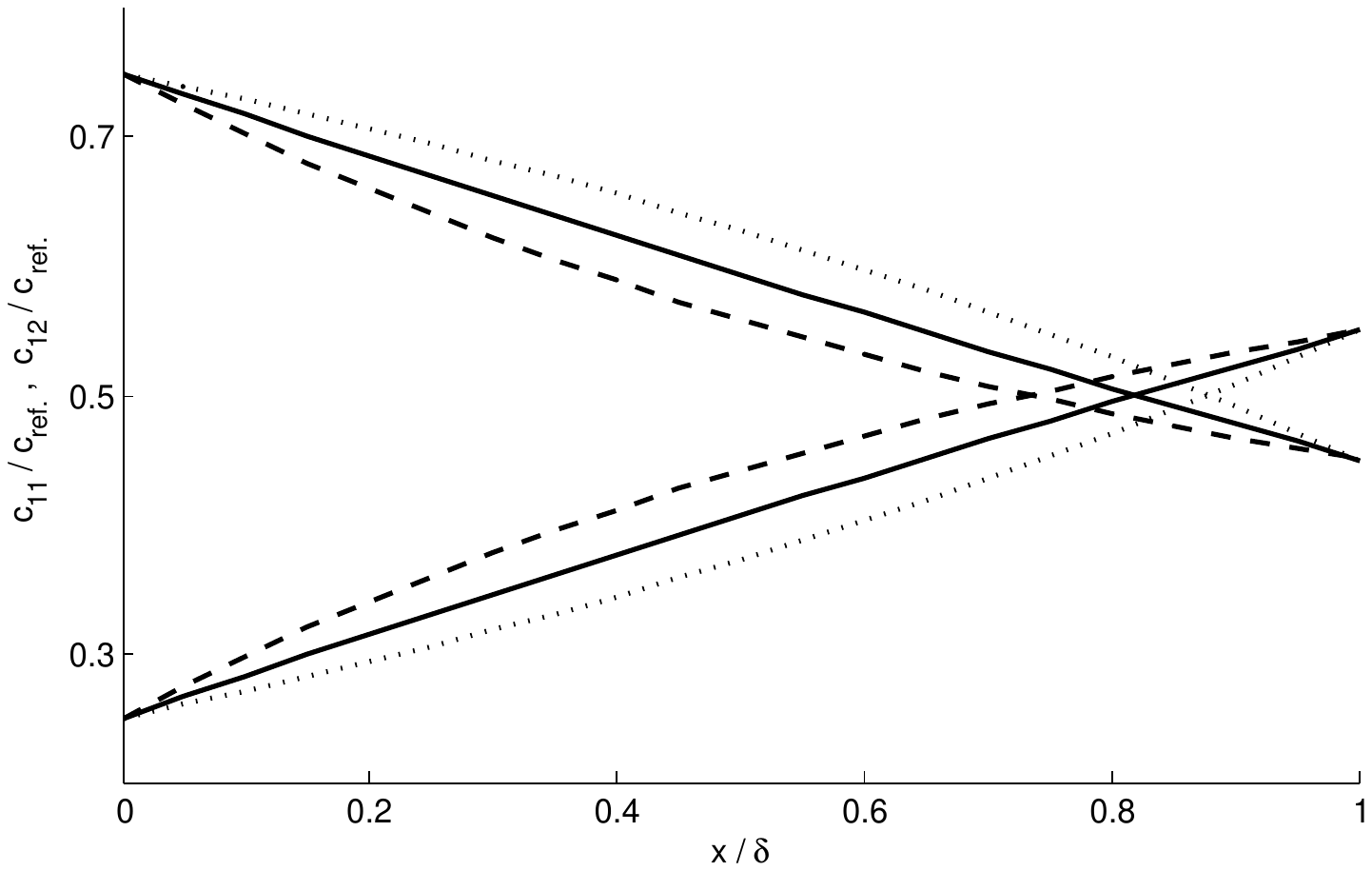}
\caption{Plots of $\hatc_{11}$ and $\hatc_{12}$ {\em v.} $\xihat$, when $\hatJ=1$ (dashed), $\hatJ=0$ (solid),  and $\hatJ=-1$ (dotted). In each case $\hatQ=-1$ and $\hatc_{11}+\hatc_{12}+\hatQ=0$ for all $\xihat$. }
\end{figure}

An important mathematical problem that remains is to establish uniqueness of the exact solutions of the corresponding
BV problems for the system of ODEs \eqref{sys1}.  For example, does the formula for $\hatE(\xihat)$
given by \eqref{GHK1}, together with the consequent formulas  for 
$\hatc_{kr}(\xihat)$ and$\hatA_{kr}$ given from \eqref{conc1soln}  and \eqref{conc1constraint2},
provide the unique solution of \eqref{sys1}  and \eqref{concBVs2}, when the boundary concentration values are also required
to  have values 
satisfying \eqref{neutrality0} and \eqref{ckrsoln}?    
 
One way to approach this problem may be to identify a higher-order ODE  satisfied by $\hatE(\xihat)$, 
together with the BCs on that ODE appropriate to a constant solution, so defining
a  BV problem for that single ODE.  A proof of uniqueness 
of the solution to {\em that} BV problem may be more easily obtained.  
If and when that is established, the forms of the concentrations and fluxes
needed to complete a solution of \eqref{sys1} will follow uniquely in a straightforward way as in the examples above.  

In the next section, such an ODE and BCs for $\hatE(\xihat)$ are obtained, for an arbitrary number of ionic 
species in an arbitrary number of valence classes. This ODE is a  generalization of that known \cite{leuchtag} 
when $\hatQ=0$, where the existence of a first integral of the system \eqref{sys1} leads to an
ODE for $\hatE$ of lower order,
and it appears to be a new equation in the context of electrodiffusion.

In the case that there is one
valence class containing any number of species, the ODE found is a form of Li\'enard equation.  For this case
uniqueness is established.

\section{Differential equation for the electric field}
Set 
\bea
\sigma_{\nu}(\xihat)=\sum_{k=1}^n\, z_k^{\nu}\,\mathcalhatC_k(\xihat)\,,\quad \alpha_{\nu}=\sum_{k=1}^n\,z_k^{\nu}\,\mathcalhatA_k\,,
\label{sigma_alphadef}
\eea
for $\nu=1\,,2\,,\dots\,,n+1$.
Now \eqref{ckODE} and \eqref{sigma_alphadef} imply that
\bea
\sigma_{\nu}\,'=\hatE\,\sigma_{\nu+1}+\alpha_{\nu}\,,
\label{sigma_recur}
\eea
and \eqref{sys1_b} and \eqref{sigma_recur} then imply by successive differentiations  that
\bea
\sigma_1=(1/\mu)\hatE\,'-\hatQ=T_1\,,\quad{\rm say}\,,
\mea\mea
\sigma_2= \frac{T_1\,'-\alpha_1}{\hatE}=T_2\,,\quad{\rm say}\,,
\label{first_couple}
\eea
and in general 
\bea
\sigma_{\nu+1}=\frac{T_{\nu}\,'-\alpha_{\nu}}{\hatE}=T_{\nu+1}\,,\quad {\rm say}\,,
\label{general}
\eea
for $\nu=1\,,2\,,\dots\,,n$.

But
\bea
\sigma_{n+1}\equiv\sigma_n\,\sum_{k}\,z_k-\sigma_{n-1}\,\sum_{k,\,l}\,'\,z_k\,z_l+\sigma_{n-2}\,\sum_{k,\,l,\,m}\,'z_k\,z_l\,z_m-
\mea\mea
\cdots +(-1)^{n-1}\sigma_1\,z_1\,z_2\,\dots\,z_n\,,
\label{sigma_expand}
\eea
where $\sum_{k,\,l}\,'$ means sum over all distinct unordered pairs of values $k$ and $l$, and so on.  It follows that
\bea
T_{n+1}=T_n\,\sum_{k}\,z_k-T_{n-1}\,\sum_{k,\,l}\,'\,z_k\,z_l+T_{n-2}\,\sum_{k,\,l,\,m}\,'z_k\,z_l\,z_m-
\mea\mea
\cdots +(-1)^{n-1}T_1\,z_1\,z_2\,\dots\,z_n\,,
\label{T_expand}
\eea
or equivalently, from \eqref{first_couple} and \eqref{general},
\bea
T_n\,'-\alpha_n=(T_{n-1}\,'-\alpha_{n-1})\,\sum_{k}\,z_k-(T_{n-2}\,'-\alpha_{n-2})\,\sum_{k,\,l}\,'\,z_k\,z_l+
\mea\mea
\cdots
+(-1)^{n-2}\,(T_1\,'-\alpha_1)\,\sum_{k,\,l,\,\dots,\,m}\,'\,z_k\,z_l\,\dots\,z_m
\mea\mea
 +(-1)^{n-1}\,\hatE\,[(1/\mu)\hatE\,'-\hatQ]\,z_1\,z_2\,\dots\,z_n\,.
\label{TODE1}
\eea
This is the autonomous ODE of order $(n+1)$ that $\hatE$ must satisfy.  Note that the last summation
in \eqref{TODE1} is over all distinct unordered $(n-1)$-tuples of values $k,\,l,\,\dots,\,m$. 
From \eqref{first_couple} and \eqref{general},
\bea
T_1=(1/\mu)\,\hatE\,'-\hatQ\,,
\quad
T_2=\frac{1}{\hatE}\,[(1/\mu)\,\hatE\,''-\alpha_1]\,,
\mea\mea
T_3=\frac{1}{\hatE\,^3}\,[(1/\mu)\,\hatE\,\hatE\,'''-(1/\mu)\,\hatE\,'\,\hatE\,''+\alpha_1\,\hatE\,'-\alpha_2\,\hatE\,^2]\,,
\label{T_forms}
\eea
so the first two ODEs are given from \eqref{T_expand} or \eqref{TODE1} by
\begin{subequations}\label{exampleODEs} 
\bea
&\,&{\bf n=1}\,:\quad \hatE\,''-z_1\,\hatE\,\hatE\,'+\mu z_1\,(\hatQ\,\hatE - \hatA_1)=0
\label{exampleODEs_a}
\\ \nonumber
\\ \nonumber
&\,&{\bf n=2}\,:\quad \hatE\,\hatE\,'''-\hatE\,'\,\hatE\,''
-(z_1+z_2)\,\hatE\,^2\,\hatE\,''
\mea\mea
&\,&\qquad\qquad +z_1z_2\,\hatE\,^3\,\hatE\,'
+\mu\,(z_1\hatA_1+z_2\hatA_2)\,\hatE\,'
\mea\mea
&\,&\qquad -\mu\,z_1\,z_2\,\hatQ\,\hatE\,^3+\mu\,z_1\,z_2\,(\hatA_1+\hatA_2)\,\hatE\,^2
=0\,. 
\label{exampleODEs_b}
\eea
\end{subequations}
The ODE \eqref{exampleODEs_a} for $n=1$ is a form of Li\'enard's equation \cite{lienard}, 
but those for higher values of $n$ are not familiar.

In earlier studies \cite{leuchtag,rogersD}, use has been made of the first integral of the system \eqref{sys1}
that exists when $\hatQ=0$, namely
\bea
\sum_{k=1}^n\,\hatc_k-\half (1/\mu)\,\hatE\,^2-\xihat\,\sum_{k=1}^n\,\hatA_k=B\quad ({\rm const.})\,,
\label{first_integral}
\eea
to obtain an ODE of order $n$ for $\hatE$.  This known $n$-th order ODE can be obtained here in a different way by 
noting that when $\hatQ=0$, \eqref{TODE1} can be integrated once to give
\bea
T_n=T_{n-1}\,\sum_{k}\,'\,z_k-T_{n-2}\,\sum_{k,\,l}\,'\,z_k\,z_l+
\mea\mea
\cdots +\half(-1)^{n-1}\,(1/\mu)\hatE\,^2\,z_1\,z_2\,\dots\,z_n
\mea\mea
+[\alpha_n-\alpha_{n-1}\,\sum_{k}\,'\,z_k+\alpha_{n-2}\,\sum_{k,\,l}\,'\,z_k\,z_l-
\mea\mea
\dots +(-1)^{n-1}\,\alpha_1\,\sum_{k,\,l,\,\dots,\,m}\,'\,z_k\,z_l\dots z_m]\,\xihat+B\,.
\label{reduced_TODE}
\eea   
Thus when $n=2$ and $z_1=-z_2=1$,  \eqref{exampleODEs_b} integrates once to give
\bea
\hatE\,''-\half\hatE\,^3-\mu\,[(\hatA_1+\hatA_2)\xihat+B]\,\hatE-\mu(\hatA_1-\hatA_2)=0\,,
\label{PII}
\eea
which is the form of PII much discussed in the literature for this case 
\cite{rubinstein,ben,zaltzman,rogersA,rogersB,thompson,amster,bracken1,bracken2} since it was first
discovered more than fifty years ago \cite{bass1}.  

The BCs associated with the ODEs for \eqref{TODE1} can be found  as follows.
From  \eqref{sigma_alphadef} and \eqref{general},
\bea
T_{\nu}(0)=\sigma_{\nu}(0)=\sum_{k=1}^n\,z_k^{\nu}\,\hatc_{k0}\,,\quad 
T_{\nu}(1)=\sigma_{\nu}(1)=\sum_{k=1}^n\,z_k^{\nu}\,\hatc_{k1}\,,
\label{EBCs1}
\eea
for $\nu=1,\,2,\,\dots,\,n$.  

Thus for $\nu=1$, 
\bea
(1/\mu)\,\hatE\,'(0)=\hatQ+\sum_{k=1}^n\,z_k\,\hatc_{k0}\,,\quad (1/\mu)\,\hatE\,'(1)=\hatQ+\sum_{k=1}^n\,z_k\,\hatc_{k1}\,,
\label{EBCs2}
\eea
for $\nu=2$,
\bea
(1/\mu)\,\hatE\,''(0)=\hatE(0)\,\sum_{k=1}^n\,z_k^2\,\hatc_{k0}+\sum_{k=1}^n\,z_k\,\hatA_k\,,
\mea\mea
(1/\mu)\,\hatE\,''(1)=\hatE(1)\,\sum_{k=1}^n\,z_k^2\,\hatc_{k1}+\sum_{k=1}^n\,z_k\,\hatA_k\,,
\label{EBCs3}
\eea
and so on.  
Then for the cases in \eqref{exampleODEs}, the BCs are
\begin{subequations}\label{exampleBCs} 
\bea
{\bf n=1}\,&:&\hatE\,'(0)=\mu\,(z_1\,\hatc_{10}+\hatQ)\,,\quad  
\hatE\,'(1)=\mu\,(z_1\,c_{11}+\hatQ)
\label{exampleBCs_a}
\\ \nonumber
\\ \nonumber
{\bf n=2}\,&:&\hatE\,'(0)=\mu\,(z_1\,\hatc_{10}+z_2\,\hatc_{20}+\hatQ)\,,
\mea\mea
&\,&\hatE\,'(1)=\mu\,(z_1\,\hatc_{11}+z_2\,\hatc_{21}+\hatQ)\,,
\label{exampleBCs_b} 
\\ \nonumber
\\ \nonumber
&\,&\hatE\,''(0)=\mu\,\hatE(0)\,(z_1^2\,\hatc_{10}+z_2^2\,\hatc_{20})+\mu\,(z_1\,\hatA_1+z_2\,\hatA_2)\,,
\mea\mea
&\,&\hatE\,''(1)=\mu\,\hatE(1)\,(z_1^2\,\hatc_{11}+z_2^2\,\hatc_{21})+\mu\,(z_1\,\hatA_1+z_2\,\hatA_2)\,.
\label{exampleBCs_c}
\eea
\end{subequations}
For general $n$ it is common to impose charge-neutrality at the boundaries as in \eqref{neutrality0},
and then the two BCs for $n=1$ in \eqref{exampleBCs_a}, and the first two BCs for $n=2$ in \eqref{exampleBCs_b}, 
simplify to
\bea
\hatE\,'(0)=0=\hatE\,'(1)\,.
\label{charge_neutralBCs2}
\eea

Note that there are $2n$ BCs in \eqref{EBCs1} although the ODE \eqref{TODE1} for $\hatE$ is only of order $(n+1)$.  
This reflects the fact that not only $\hatE$, but also the $\hatA_k$ appearing in the ODE and in the BCs, are unknowns 
to be determined as part of a solution of the system of equations \eqref{sys1}. 

 This adds significantly to the difficulty of solving the ODE   numerically with the BCs \eqref{EBCs1}, 
 even for small values of $n$.  For example,  in the case $n=2$ with $\hatQ=0$, and  after \eqref{TODE1} has been
 integrated once to give the
 Painlev\'e equation \eqref{PII},    a direct numerical attack  proves quite difficult \cite{amster}.  
 A way of avoiding these difficulties is to solve the first order system defined by \eqref{sys1_b} and \eqref{ckODE}
 from which \eqref{TODE1} and \eqref{EBCs1} are derived, after augmenting it with ODEs
 expressing the constancy of the $\hatA_{k}$ in order to construct an enlarged system 
 for which the number of BCs matches the 
 number of dependent variables.  
 
 To illustrate, consider the case $n=2$ with $z_1=-z_2=1$, $m_1=m_2=1$,  and the 
 ODE \eqref{exampleODEs_b} with BCs \eqref{exampleBCs_b} and \eqref{exampleBCs_c}, which  follow from the system
 \bea
 \hatc_1\,'(\xihat)=\hatE(\xihat)\,\hatc_1(\xihat)+\hatA_1\,,\quad \hatc_2\,'(\xihat)=
 -\hatE(\xihat)\,\hatc_2(\xihat)+\hatA_2\,,
 \mea\mea
 \hatE\,'(\xihat)=\mu\left(\hatc_1(\xihat)-\hatc_2(\xihat)+\hatQ\right)\qquad\qquad\qquad
 \label{example_sys1}
 \eea
with BCs $\hatc_k(0)=\hatc_{k0}$,  $\hatc_k(1)=\hatc_{k1}$, for $k=1\,,2$, together with a prescribed value for
$\hatJ=-\hatD_1\,\hatA_1+\hatD_2\,\hatA_2$, making five constraints in total.  
Note that there are five unknowns to be determined, namely
$\hatE(\xihat)$, the two $\hatc_k(\xihat)$ and the two constants $\hatA_k$.   The suggestion is to replace 
\eqref{example_sys1} and the five constraints by the augmented system of five ODEs
\bea
\hatc_1\,'(\xihat)=\hatE(\xihat)\,\hatc_1(\xihat)+\hatA_1(\xihat)\,,\quad \hatc_2\,'(\xihat)=
 -\hatE(\xihat)\,\hatc_2(\xihat)+\hatA_2(\xihat)\,,
 \mea\mea
 \hatE\,'(\xihat)=\mu\left(\hatc_1(\xihat)-\hatc_2(\xihat)+\hatQ\right)\,,\quad \hatA_1\,'(\xihat)=0\,,\quad  \hatA_2\,'(\xihat)=0\,,
 \label{example_sys2}
 \eea
with five associated  BCs
\bea
\hatc_1(0)=\hatc_{10}\,,\quad \hatc_1(1)=\hatc_{11}\,,\quad \hatc_2(0)=\hatc_{20}\,,\quad\hatc_2(1)=\hatc_{21}\,,
\mea\mea
-\hatD_1\,\hatA_1(0) + \hatD_2\,\hatA_2(0)=\hatJ\,.\qquad\qquad\qquad
\label{example_BCs2}
\eea
Now the problem is in a standard form that is easily handled by
a  standard ODE solver \cite{matlab}.

This technique was used to obtain the solution $\hatE(\xihat)$ of \eqref{exampleODEs_b},  
\eqref{exampleBCs_b} and \eqref{exampleBCs_c}, together with the associated 
values of $\hatA_1$ and $\hatA_2$, for a variety of 
values of $\hatQ$ , in order to detect any significant differences in behaviour of the solutions
between cases with $\hatQ\ne 0$ and the case $\hatQ=0$, which integrates once to the Painlev\'e equation \eqref{PII}.
Fig. 5 shows plots of $\hatE$ {\em v.} $\xihat$, for $\hatQ=1,\,0.5,\,0,\,-0.5,\,-1$ from top to bottom, in an example with
$\mu=0.3$ and $\hatJ=-0.4$. Dimensionless diffusion coefficients were taken to be $\hatD_1=0.4$, $\hatD_2=0.6$ , 
as appropriate for Na$^+$ and Cl$^-$ \cite{unsw}.
Charge-neutral BCs were adopted in each case, with $\hatc_{10}=1.5=\hatc_{20}-\hatQ$,
$\hatc_{11}=1.1=\hatc_{21}-\hatQ$. The corresponding (constant) values of $(\hatA_1,\,\hatA_2)$ obtained were
$(-0.18,\,-0.79),\,(-0.14,\,-0.76)$,
$(-0.08,\,-0.72),\,(0.01,\,-0.66),\,(0.19,\,-0.57)$, each consistent with the last of \eqref{example_BCs2}.

\begin{figure}[ht]
\centering
\includegraphics[width=5in]
{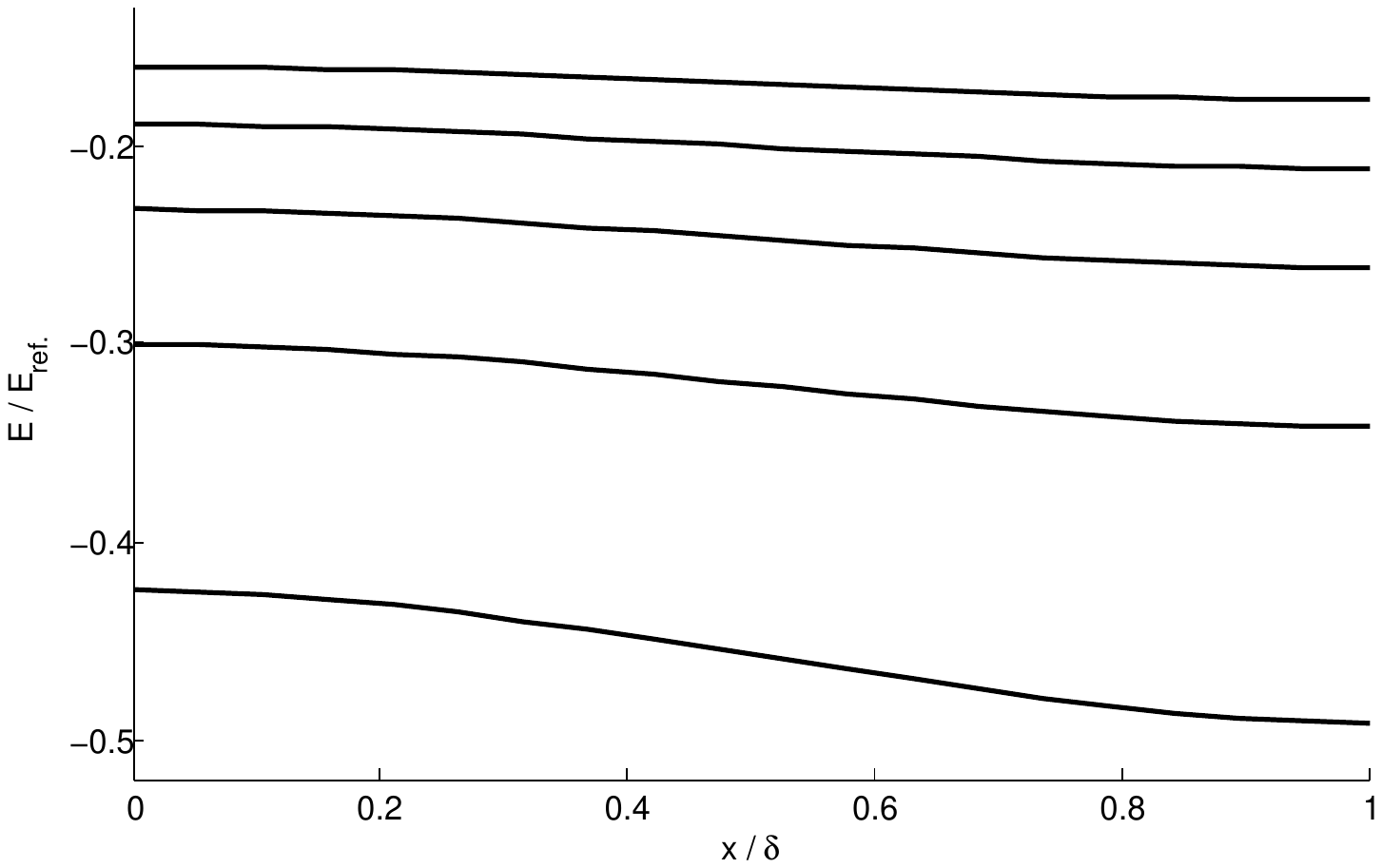}
\caption{Plots of $\hatE(\xihat)$ {\em v.} $\xihat$ for $\hatQ=1$ (top), $\hatQ=0.5$, $\hatQ=0$, $\hatQ=-0.5$,
down to  $\hatQ=-1$ (bottom). }
\end{figure}

The plots show no marked difference in character between the cases, and suggest a smooth transition from 
$\hatQ\ne 0$ to $\hatQ=0$.  The nature of singularities that can appear in solutions of PII is well-known 
and widely-documented \cite{painleve,clarkson}, but no sign of
any singularities occurs in numerical studies of \eqref{PII} on the finite $\xihat$- interval with 
charge-neutral BCs \cite{bracken1}, 
and it seems likely that the same is true for
\eqref{exampleODEs_b} with $\hatQ\ne 0$.  On the other hand, singularities are known to arise for \eqref{PII} with 
more general BCs \cite{bracken2},  and it would be interesting to examine \eqref{exampleODEs_b} both analytically 
and numerically in such cases, and also
possibly on unbounded regions of the $\xihat$-axis, 
or even on the complex $\xihat$-plane, to identify any new behaviours.

It is not clear for example that when $\hatQ\ne 0$,  \eqref{exampleODEs_b} still has the Painlev\'e property, namely
that its general solution is free of branch points whose location in the complex plane is dependent on one or more 
constants of integration. The ODE does not appear explicitly in Chazy's classification of third order ODEs with this property 
\cite{chazy,cosgrove}, but it remains possible that it could be reduced to such an equation by suitable changes of variables.
 It can be said that \eqref{exampleODEs_b} is 
a generalization of PII in the sense that its
solutions must depend on the parameter $\hatQ$ in such a way that, as $\hatQ\to 0$, they 
can be expressed in terms of Painlev\'e transcendents, but no stronger claim can be made without further analysis. 
Studies of the singularity properties of  \eqref{exampleODEs_b}, and more generally of the equations \eqref{TODE1}, 
are beyond the scope of the present work.

\section{Li\'enard ODE and uniqueness of constant field solution}
   
When $n=1$, there is one class containing $m_1$ species with valence $z_1$. 
The electric field $\hatE$ satisfies the Li\'enard equation \eqref{exampleODEs_a}, and assuming 
charge-neutrality on the boundaries as in \eqref{neutrality0}, it also satisfies the  BCs \eqref{charge_neutralBCs2}. 
 
It is obvious by inspection that the BV problem defined by \eqref{exampleODEs_a} and \eqref{charge_neutralBCs2} 
admits a constant solution
\bea
\hatE(\xihat)=\hatA_1/\hatQ=\hatES\,,\!\!\quad{\rm say},
\label{ESdef}
\eea
bearing in mind that the constant  $\hatA_1$ is as yet undetermined.   It follows from \eqref{sys1_b} that with $\hatE$ constant,
\bea
z_1\,\hatc_1(\xihat)+\hatQ=0\,,\quad 0<\xihat<1\,.
\label{neutrality3} 
\eea
in other words, charge-neutrality holds throughout the membrane, and not only on the boundaries, as was assumed in 
\eqref{neutrality0} and hence in \eqref{charge_neutralBCs2}.  

This is a surprise. It is not obvious that the assumption of  
charge-neutrality at the boundaries should imply charge-neutrality throughout the membrane, and this raises the 
question of the uniqueness of the constant 
solution \eqref{ESdef} to the problem posed by \eqref{exampleODEs_a} and \eqref{charge_neutralBCs2} 
(for  any particular value of the constant
$\hatA_1$ ).  Perhaps there are other, non-constant solutions
which, according to \eqref{sys1_b},  would not imply charge-neutrality?

However,  \eqref{ESdef} is indeed the unique solution of \eqref{exampleODEs_a} 
and \eqref{charge_neutralBCs2}, 
so that \eqref{ESdef} and 
\eqref{neutrality3} do both follow from the assumption \eqref{concBVs2_b}.  
This can be seen as follows.  

Note firstly that \eqref{neutrality3} cannot be satisfied unless $\hatQ$ is nonzero, with opposite sign to $z_1$.  
Assuming this is so, then
\eqref{exampleODEs_a} can be rewritten as
\bea
\hatE\,''(\xihat)-z_1\,\hatE(\xihat)\,\hatE\,'(\xihat)-\alpha^2[\hatE(\xihat)-\hatES]=0\,,
\label{DE}
\eea
where $\alpha^2=-\mu\,\hatQ\,z_1  >0$, with $\hatES$ as in \eqref{ESdef}.  The problem is to show the uniqueness  of the solution
\eqref{ESdef} of this ODE on $0<\xihat<1$ with the BCs \eqref{charge_neutralBCs2}.
To be precise, by a solution here is meant a function $\hatE(\xihat)$ with smooth second  derivative for  $0<\xihat<1$,
satisfying \eqref{DE} on that interval, and also
\bea
\hatE\,'(0+)=0=\hatE\,'(1-)\,.
\label{preciseBCs}
\eea
 \vs\ni
{\bf (A)} Note firstly that no solution can have a maximum at any $\xihat_m$ with $0<\xihat_m<1$
and  with $\hatE(\xihat_m)>\hatES$.  
For at such a point,  $\hatE\,''(\xihat_m)\leq 0$,
$\hatE\,'(\xihat_m)=0$ and $[\hatE(\xihat_m)-\hatES)]>0$, contradicting \eqref{DE}.  
\vs\ni
Similarly,  no solution can have a minimum at any $\xihat_m$ with $0<\xihat_m<1$ 
and with $\hatE(\xihat_m)<\hatES$.  
\vs\ni
{\bf (B)} Now suppose there exists a solution with $\hatE(0+)=\hatE_0>\hatES$.  
Then \eqref{DE} implies that $\hatE\,''(0+)>0$, so that the graph of $\hatE(\xihat)$ is concave 
upwards at $\xihat=0+$.  
\vs\ni
{\bf (a)} Suppose in addition to {\bf (B)} that $\hatE(1-)=\hatE_1>\hatES$. Then the graph of 
$\hatE(\xihat)$ is also concave upwards at $\xihat=1-$, and it follows that 
$\hatE(\xihat)$ must have a maximum at some $\xihat_m$ with $0<\xihat_m<1$ and with $\hatE(\xihat_m)>\hatES$, 
implying a contradiction by {\bf (A)}. 
\vs\ni
{\bf (b)} Suppose instead in addition to {\bf (B)} that $\hatE(1-)=\hatE_1\leq \hatES$.  
Then again it follows that $\hatE(\xihat)$ must have a maximum at some $\xihat_m$ with
$0<\xihat_m<1$ and with $\hatE(\xihat_m)>\hatES$, again implying a contradiction by {\bf (A)}. 
\vs\ni
Thus there is no solution with  $\hatE(0+)=\hatE_0>\hatES$.  
Similarly, there is no solution with $\hatE(0+)=\hatE_0<\hatES$.  
\vs\ni
Then, similarly, there is 
no solution with $\hatE(1-)=\hatE_1\neq \hatES$. 
\vs\ni
{\bf (C)} Suppose instead of {\bf (B)} that there exists a solution with 
$\hatE(0+)=\hatES=\hatE(1-)$, 
but with $\hatE(\xihat)\not\equiv \hatES$. Then either $\hatE(\xihat)>\hatES$ or
$\hatE(\xihat)<\hatES$ at some points with
$0<\xihat<1$ (or both).  In the first case,  $\hatE(\xihat)$ must have a maximum at some $\xihat_m$ with
$0<\xihat_m<1$ and with $\hatE(\xihat_m)>\hatES$, 
while in the second case it must have a minimum at some $\xihat_m$ with
$0<\xihat_m<1$ and with $\hatE(\xihat_m)<\hatES$.   (In the third case, it must have both).  In any case there is a contradiction, 
by {\bf (A)}. 
\vs\ni
Thus there is no solution of \eqref{DE} and \eqref{preciseBCs} except for the constant solution \eqref{ESdef}.

In any attempt to generalize these arguments to  cases with $n\ge 2$,  more complicated BCs such as those in 
\eqref{exampleBCs_c} have to be considered, involving the unknowns $\alpha_k$ as well as the boundary values of
the concentrations $\hatc_k$.  
It is still unclear if such complications can be overcome so that a proof of uniqueness can be found in such cases.

\section{Concluding remarks}
Goldman-Hodgkin-Katz formulas are extensively used, especially in physiological settings, to obtain 
estimates of voltage differences across liquid junctions.  The electric field in the junction is assumed 
to be constant, or approximately so, in the derivation of these formulas, which express the field and hence the voltage difference
in terms of BVs of ionic concentrations when the current density across the junction vanishes. 

A different approach has been taken above, by explicitly allowing for the presence
of a uniform distribution of fixed charges, thus allowing a greater richness of {\em exact} constant field solutions and corresponding 
BVs of the concentrations to be determined. Once again  formulas of GHK type are obtained in the case of vanishing current density.   
In one case this takes a similar form to the usual formula, as in \eqref{GHK7},
in other cases they appear to be new.  
But in all cases they are exact, and not derived as approximations following from  an assumption 
that the junction width is small compared with the ``internal Debye length" as in  \eqref{debye}.
The price paid to obtain exactness is that the formulas now hold only for the particular BVs of the concentrations.

The importance of the mathematics of electrodiffusion stems from its numerous significant applications
in the physical and biological sciences.  The models involved are nonlinear, and the classical questions of 
existence and uniqueness of solutions have barely been considered in the literature.   
It is entirely reasonable to suppose, on the basis of numerical experiments,
that unique solutions exist for  two-point BV problems for the system of coupled ODEs \eqref{sys} 
with  prescribed current density \eqref{curr} and BCs \eqref{concBVs1}, when these BCs are charge-neutral.  
Apart from the explicit appearance of
exact constant field solutions as
described above, however,  existence has been proved more generally only 
in the case $n=2$,  $m_1=m_2=1$, $z_1=-z_2=1$ and $Q=0$, and uniqueness not even then.  
The proof given above of uniqueness of exact constant field solutions in the case $n=1$, $m_1$ arbitrary, $Q\ne 0$
is a modest step towards
remedying this deficiency, but the uniqueness problem remains unsolved even for exact constant field solutions, when $n>1$.

\end{document}